\documentclass[runningheads]{llncs}

\usepackage[utf8]{inputenc}
\usepackage{xcolor}
\usepackage{amsfonts}
\usepackage{amssymb}
\usepackage{comment}
\usepackage{mathtools}
\usepackage{thmtools}
\usepackage{cleveref}
\usepackage[ruled,vlined,onelanguage,noend]{algorithm2e}
\usepackage[format=plain, labelfont=bf, labelsep=none]{caption}
\DeclareCaptionLabelSeparator{none}{ }
\usepackage{float}
\usepackage{enumerate}
 \newcommand{\alain}[1]{}
\newcommand{\kp}{{k^\perp}}
\newcommand{\OO}{\mathcal{O}}
\newcommand{\CC}{\mathcal{C}}
\newcommand{\Cp}{\CC^\perp}

\newcommand{\DD}{\mathcal{D}}
\newcommand{\Dp}{\DD^\perp}

\newcommand{\PP}{\mathbb{P}}
\newcommand{\EE}{\mathbb{E}}
\newcommand{\ZZ}{\mathbb{Z}}
\newcommand{\FF}{\mathbb{F}}
\newcommand{\Fq}{\FF_q}
\newcommand{\FQ}{\FF_{q'}}

\newcommand{\Fqnn}{\Fq^{n\times n}}
\newcommand{\Fqmn}{\Fq^{m\times n}}
\newcommand{\Fqmm}{\Fq^{m\times m}}
\newcommand{\MCE}{\mathsf{MCE}}
\newcommand{\CMCE}{\mathsf{CMCE}}
\newcommand{\TFE}{\mathsf{TFE}}
\newcommand{\ATFE}{\mathsf{ATFE}}
\renewcommand{\leq}{\leqslant}

\newcommand{\eqdef}{\stackrel{\text{def}}{=}}
\newcommand{\Sep}{\mathrm{Sep}_{q,m}}

\DeclareMathOperator{\Span}{Span}
\DeclareMathOperator{\Dict}{Dict}
\DeclareMathOperator{\GL}{GL}
\DeclareMathOperator{\Gr}{Gr}
\newcommand{\GLnq}{\GL_n(\Fq)}
\newcommand{\GLmq}{\GL_m(\Fq)}
\DeclareMathOperator{\Tr}{Tr}
\renewcommand{\Pr}{{\rm Pr}}

\newcommand{\qbin}{\genfrac{[}{]}{0pt}{}}
\newcommand{\CL}[1]{}
\newcommand{\pA}[1]{}

\spnewtheorem{assumption}{Assumption}{\bfseries}{\itshape}
 
\def\arXiv{2}

\title{Highway to Hull: \\An Algorithm for Solving the General Matrix Code Equivalence Problem}
\titlerunning{Highway to Hull}
\author{Alain Couvreur\inst{1,2} \and Christophe Levrat\inst{1,2}}
\institute{Inria \and Laboratoire LIX,\\ École Polytechnique,\\ Institut Polytechnique de Paris,\\ France\\
\email{\{alain.couvreur, christophe.levrat\}@inria.fr}}
\date{}
\begin{document}

\maketitle

\begin{abstract}
  The matrix code equivalence problem consists, given two matrix spaces $\mathcal{C},\mathcal{D} \subset \Fq^{m\times n}$ of dimension $k$, in finding invertible matrices $P\in\mathrm{GL}_m(\mathbb{F}_q)$ and $Q\in\mathrm{GL}_n(\mathbb{F}_q)$ such that $\mathcal{D}=P\mathcal{C} Q^{-1}$. Recent signature schemes such as MEDS and ALTEQ relate their security to the hardness of this problem. Recent works by Narayanan, Qiao and Tang on the one hand and by Ran and Samardjiska on the other hand tackle this problem. The former is restricted to the ``cubic'' case $k = m =n$ and succeeds in $\widetilde{\mathcal{O}}(q^{\frac k 2})$ operations. The latter is an algebraic attack on the general problem whose complexity is not fully understood and which succeeds only on $\mathcal{O}(1/q)$ instances. We present a novel algorithm which solves the problem in the general case.  Our approach consists in reducing the problem to the matrix code conjugacy problem, \emph{i.e.} the case $P=Q$. For the latter problem, similarly to the permutation code equivalence problem in Hamming metric, a natural invariant based on the \emph{Hull} of the code can be used. Next, the equivalence of codes can be deduced using a usual list collision argument.  For $k=m=n$, our algorithm achieves the same time complexity as Narayanan \emph{et al.} but with a lower space complexity. Moreover, ours extends to a much broader range of parameters.
\end{abstract}

\section*{Introduction}
In the last decades, so-called \emph{equivalence problems}
have frequently been used for cryptographic applications. The first examples probably
come from multivariate cryptography with the Matsumoto-Imai \cite{MI88} or HFE \cite{P96} schemes,
whose security relies on the hardness of the polynomial isomorphism
problem: deciding whether two spaces of polynomials are equivalent
with respect to a linear or affine change of variables.

In recent years, we observed an intensification of this trend but
also a diversification of the equivalence problems used for
cryptography.  In particular, in NIST's recent on-ramp call for
signature\footnote{\url{https://csrc.nist.gov/Projects/pqc-dig-sig/round-1-additional-signatures}},
many signature schemes involve equivalence problems which are not the
polynomial isomorphism one. For instance, \emph{Hawk}'s \cite{Hawk}
security rests among others on the hardness of the Lattice Isomorphism
Problem (LIP), LESS \cite{LESS} rests on the monomial equivalence of
Hamming metric codes, MEDS \cite{MEDS} on the matrix code equivalence
and ALTEQ \cite{ALTEQ} on the equivalence of alternate trilinear
forms.  The latter problem (equivalence of alternate trilinear forms)
is in fact a sub-case of the former: the matrix code equivalence
problem, which is the purpose of the present article. The two problems are actually proven to be polynomially equivalent
\cite[Prop.~8.3]{grochow2023complexity}.

Given two matrix spaces $\CC, \DD \subset \Fq^{m \times n}$, the \emph{matrix code equivalence problem} consists in deciding whether there exists $P \in \GL_m (\Fq)$ and $Q \in \GL_n (\Fq)$
such that $\DD = P \CC Q^{-1}$. The search version of the problem asks to return, if exists, a pair $(P,Q)$ providing the equivalence. Even though its use for cryptography is rather new, this problem
has been known for a long time in algebraic complexity theory where it is usually formulated
in an equivalent way as the \emph{$3$--tensor isomorphism} problem. This problem is assumed to be hard and is in particular known to be at least as hard as the monomial code equivalence problem (see \cite{grochow2023complexity} or \cite{CDG20}).

\subsection*{Our contribution} In this article, we present a new algorithm for solving the
matrix code equivalence problem, or equivalently the
$3$--tensor isomorphism problem. Given equivalent $k$--dimensional $m \times n$ matrix spaces with entries in $\Fq$, we are able to solve the search equivalence problem in  \[\widetilde{\OO}(q^{\max(\frac{k}{2}, k-m+2)})\] operations in $\Fq$. Note in particular that in the specific case $k = m =n$
which is the one that is used in the parameters of MEDS and ALTEQ, we achieve the time complexity
$\widetilde \OO(q^{\frac k  2})$ which is the one achieved by Narayanan, Qiao and Tang \cite{nqt}.
However,
\begin{enumerate}
    \item our algorithm rests on completely different invariants;
    \item in the specific case $k=m=n$, the space complexity of our algorithm is $\OO(nq^{\frac{n}{2}-1})$, which is is smaller than that of \cite{nqt} by a factor $n^2$;
    \item our result does not require the parameters $k,m,n$ to be equal, while this is necessary for Narayanan \emph{et. al.}'s algorithm to run.
    
    Note that, if the equivalence of alternate trilinear forms
    problem on which ALTEQ is built requires by design to have $k = m =n$, there is no need to instantiate MEDS with such a constraint
    on $k, m, n$. It turns out that MEDS' proposed parameters \cite{MEDS} satisfy this condition making them vulnerable to
    Narayanan \emph{et. al.}'s attack, but MEDS' designers could have easily circumvented the aforementioned attack just by breaking the symmetry on the parameters $k,m,n$. Still, the algorithm introduced in the present article attacks a much broader range of triples
    $k,m,n$.
\end{enumerate}
A specificity of our algorithm is that, taking its inspiration from the Hamming metric
counterpart of the code equivalence problem problem and Sendrier's famous \emph{support splitting algorithm} \cite{S00},
we use the \emph{Hull} of the code, \emph{i.e.} its intersection with its orthogonal space w.r.t some given bilinear form.

\subsection*{Related works}
The schemes MEDS and ALTEQ \cite{MEDS,ALTEQ} were both submitted to NIST's on-ramp call for digital signatures. Before, ALTEQ's and MEDS' specifications were respectively presented in the articles \cite{ALTEQpaper} and \cite{MEDSpaper}.
In \cite{B23}, Beullens describes a new algorithm
solving the trilinear form equivalence problem, harming the proposed parameters for ALTEQ. More recently, Narayanan, Qiao and Tang \cite{nqt} presented an algorithm
solving the same problem but also the matrix code equivalence problem in the case of $k$--dimensional spaces of $k \times k$ matrices. Their approach combines a collision list argument with a nice algebraic invariant and achieves a complexity in $\widetilde{\OO} (q^{\frac k 2})$. Finally, Ran and Samardjiska \cite{ran24} designed an algorithm for the 3-tensor isomorphism problem which looks for triangles in tensor graphs. Such triangles exist in roughly $1/q$ of all instances of the problem. In these instances and for current parameters of MEDS and ALTEQ, their algorithm provides a speedup compared to all previous works.

\if\arXiv{2}
\subsection*{Outline of the article}
In \Cref{sec:codequiv}, we introduce the matrix code equivalence problem as well as some related problems and algorithms solving them. In \Cref{sec:techover}, we first state some key observations at the heart of our algorithm, then give a brief presentation of the algorithm as well as its complexity. The algorithm itself can be divided into three steps. The crucial one is the reduction to a specific instance of the matrix code conjugacy problem, where conjugate one-dimensional subspaces inside the two codes are given. This reduction is presented in \Cref{sec:reduction}. We then solve the conjugacy problem in question using an algorithm presented in \Cref{sec:find}. Finally, we explain in \Cref{sec:recoverQ} how to deduce a solution to the initial matrix code equivalence problem from the solution to the conjugacy problem found in the previous step.
\fi

\if\arXiv1
\subsection*{Acknowledgements}
The authors express their deep gratitude to Matthieu Lequesne who suggested the title. They also thank the anonymous referees for their very helpful comments.
\if\arXiv2

\fi
The second author is funded by Inria and the French Cybersecurity Agency (ANSSI). This work benefited from the financial
support of the French government and the \emph{Agence Nationale de la Recherche (ANR)} through the \emph{Plan France 2030} via the project ANR-22-PETQ-0008. The authors are part of a collaborative research project \emph{Barracuda} with reference ANR-21-CE39-0009-BARRACUDA. The first author is partially funded by Horizon-Europe MSCA-DN project \emph{Encode}.

\fi
 
\section{The matrix code equivalence problem}
\label{sec:codequiv}

\begin{definition} Let $m,n,k$ be positive integers. 
An $m\times n$ matrix code of dimension $k$ is a $k$-dimensional $\Fq$-linear subspace of $\Fqmn$.
\end{definition}

\begin{definition}[Matrix code equivalence problem]\label{def:mce} Let $m,n,k$ be positive integers. 
Consider two $k$-dimensional linear subspaces $\CC,\DD$ of $\Fqmn$. 
The matrix code equivalence problem $\MCE_{m,n,k}(\CC,\DD)$ consists in finding (if exist) matrices $P\in\GL_m(\Fq),Q\in\GL_n(\Fq)$ such that \[\DD=P\CC Q^{-1}.\]
When $m=n=k$, we call it the \emph{cubic matrix code equivalence problem}: $\CMCE_{n}(\CC,\DD)$.
\end{definition}

\begin{remark} We may suppose that $m\leq n$. Indeed, if $\DD=P\CC Q^{-1}$ then \[ \DD^\top =(Q^{-1})^\top\CC^\top P^\top\] so any algorithm solving the case $m\leq n$ can also be used, after transposing the whole problem, to solve the case $n\leq m$. In the remainder of this article, we will always suppose that $m\leq n$. Moreover, we may switch $\CC$ and its dual $\Cp$ in order to have $\dim(\Cp)\leqslant \dim(\CC)$. Indeed, if $\DD=P\CC Q^{-1}$ then \[ \Dp =(P^{-1})^\top \Cp Q^\top.\] Hence, the case where $mn-k=m=n$ may be reduced to an instance of $\CMCE$.
\end{remark}

The $\CMCE$ problem is notably the basis of the former NIST signature scheme candidate MEDS. 
A polynomial-time equivalent problem \cite{algebraic_trilin}, the \emph{alternating trilinear form equivalence problem}, underpins the former NIST signature candidate ALTEQ. 
An attack against these problems was recently described by Naranayan, Qiao and Tang in \cite{nqt}.

\subsection{Related problems}\label{subsec:related}

\subsubsection{The trilinear forms equivalence problem} 

\begin{definition}[Trilinear Form Equivalence Problem ($\TFE$)] The \emph{trilinear forms equivalence} problem $\TFE_{m,n,k}$ is the following. Given two trilinear forms $f,g\colon \Fq^m\times\Fq^m\times\Fq^k\to \Fq$, find three matrices $(P,Q,R)\in\GLmq\times\GLnq\times\GL_k(\Fq)$ such that for any $x,y,z\in\Fq^m\times\Fq^n\times\Fq^k$, \[f(Px,Qy,Rz)=g(x,y,z).\]
\end{definition}

The following well-known result shows the the matrix code equivalence problem reduces to the trilinear forms equivalence problem with the same parameters.
\begin{lemma}The problem $\MCE_{m,n,k}$ admits a (deterministic) polynomial-time reduction to $\TFE_{m,n,k}$. 
\end{lemma}

\if\arXiv2
\begin{proof}Let $(\CC,\DD)$ be an instance of $\MCE_{m,n,k}$. Denote by $(C_1,\dots,C_k)$ a basis of $\CC$ and by $(D_1,\dots,D_k)$ a basis of $\DD$.  We may define the trilinear forms
\begin{align*}
    f\colon (x,y,z)&\mapsto \sum_{i,j,\ell}{(C_k)}_{ij}x_iy_jz_\ell \\
    g\colon (x,y,z)&\mapsto \sum_{i,j,\ell}{(D_k)}_{ij}x_iy_jz_\ell.
\end{align*}
If $g(x,y,z)=f(Px,Qy,Rz)$ for all $(x,y,z)\in\Fq^m\times\Fq^n\times\Fq^k$, then straightforward computations show that $\DD=P^\top \CC Q^\top $, and the element $R_{ij}$ is the $i$-th coordinate of $D_j$ when expressed in the basis $(P^\top C_1Q^\top ,\dots,P^\top C_kQ^\top )$ of $\DD$. \alain{double check}\qed
\end{proof}

\begin{remark}
  Even if the two aforementioned problems are actually polynomially
  equivalent (see \cite{grochow2023complexity}), the converse of this construction
  does not directly yield a deterministic polynomial-time reduction of
  $\TFE_{m,n,k}$ to $\MCE_{m,n,k}$. Indeed, let $(f,g)$ be an instance
  of $\TFE_{m,n,k}$. We may write \begin{align*}f&\colon
    (x,y,z)\mapsto \sum_{i,j,\ell} c_{ijk}x_iy_jz_\ell \\ g&\colon
    (x,y,z)\mapsto \sum_{i,j,\ell} d_{ijk}x_iy_jz_\ell.\end{align*} For
  $r\in\{1\dots k\}$, construct the matrices $C_r=(c_{ijr})_{i,j}$ and
  $D_r=(d_{ijr})_{i,j}$.  Now, it is not guaranteed that the codes
  $\CC=\Span(C_1,\dots,C_k)$ and $\DD=\Span(D_1,\dots,D_k)$ are
  $k$-dimensional. However, if they are, this is indeed an instance of
  $\MCE_{m,n,k}$ and if we find a solution $(P,Q)$ such that $\DD=P\CC Q$, we can immediately solve this instance of $\TFE_{m,n,k}$. Indeed, consider the matrix $R=(r_{ij})\in\GL_k(\Fq)$ where $r_{ij}$ is
  $i$-th coordinate of $D_j$ when expressed in the basis $(PC_1Q,\dots,PC_kQ)$ of
  $\DD$. Then, we have $f(P^\top x, Q^\top y ,R z)=g(x,y,z)$ for all
  $(x,y,z)$. In practice, given a random trilinear form, the matrices
  $C_i$ are random elements of $\Fq^{m\times n}$, and they are very
  likely to be linearly independent.
\end{remark}
\fi

\subsubsection{Alternate trilinear form equivalence problem}
A sub-case of the trilinear form equivalence problem that has been
considered for the design of ALTEQ is the \emph{alternate trilinear
forms equivalence problem} $\ATFE$. An \emph{alternate} trilinear
form is a trilinear form $f : \Fq^n \times \Fq^n \times \Fq^n$
such that for any $x \in \Fq^n$,
\[
f(x,x,\cdot) \equiv f(x,\cdot, x) \equiv f(\cdot, x,x) \equiv 0.
\]
This is equivalent to the fact that, given any permutation $\sigma \in \mathfrak{S}_3$ (the group of permutation on $3$ letters),
and any triple $(x_1,x_2,x_3) \in (\Fq^n)^3$,
\[
f(x_{\sigma (1)}, x_{\sigma (2)}, x_{\sigma (3)}) = \varepsilon (\sigma)
f(x_1, x_2, x_3),
\]
where $\varepsilon (\sigma)$ denotes the signature of the permutation
$\sigma$.
This definition leads to the following problem.

\begin{definition}[Alternate Trilinear Form Equivalence Problem ($\ATFE$)] The alternate trilinear forms equivalence problem $\ATFE_{m,n,k}$ is the following. Given two alternate trilinear forms $f,g\colon \Fq^m\times\Fq^m\times\Fq^k\to \Fq$, find a matrix $P\in\GLnq$ such that for any $x,y,z\in\Fq^m\times\Fq^n\times\Fq^k$, \[f(Px,Py,Pz)=g(x,y,z).\]
\end{definition}

\subsubsection*{$3$--tensor isomorphism.}
On the tensor product $\Fq^m \otimes \Fq^n \otimes \Fq^k$,
there is a natural action of $\GL_m (\Fq) \times \GL_n(\Fq) \times \GL_{k} (\Fq)$, and the $3$--tensor isomorphism problem consists
in deciding whether two tensors $T_1, T_2 \in \Fq^m \otimes \Fq^n \otimes \Fq^k$ are in the same orbit with respect to the aforementioned group action.
\if\arXiv1
The $3$--tensor isomorphism problem is well-known to be equivalent
to the matrix code equivalence one. See \cite[Fig.~2]{grochow2023complexity}
for further details on the connections between various equivalence problems.
\fi

\if\arXiv2
The equivalence between the matrix code equivalence problem and the
$3$--tensor isomorphism one, is very explicit. Given two tensors, one can consider the matrix subspaces of $\Fq^{m}\times \Fq^n$
spanned by their ``slices'' and the tensors are isomorphic if and only if the corresponding matrix spaces are equivalent.
Conversely, given two matrix spaces, one can take a basis for each one,
and stack elements of a basis in order to create a $3$--tensor. Then,
the matrix spaces will be equivalent if and only if the corresponding
$3$--tensors are isomorphic.
\fi

\begin{remark}
Note that the terminology of \emph{cubic matrix code equivalence problem} introduced in Definition~\ref{def:mce} refers to the corresponding tensors, which will be $n \times n \times n$, \emph{i.e.}
\emph{cubic tensors}.
\end{remark}

\if\arXiv2
Similarly, the equivalence between the $3$--tensor isomorphism and
the equivalence of trilinear forms, can be made explicit since a trilinear form is encoded by a 3-tensor $T\in \Fq^m\otimes \Fq^n\otimes \Fq^k$. The equivalence of the problems mentioned in this section and more is summarized in \cite[Figure 2]{grochow2023complexity}.

Finally, $\ATFE$ can be reformulated in terms of the equivalence 
of \emph{alternate tensors} which are tensors $T \in\Fq^n \otimes \Fq^n
\otimes \Fq^n$ such that for any $\sigma \in \mathfrak{S}_3$,
\[
    \sigma(T) = \varepsilon (\sigma) \cdot T,
\]
where $\sigma (T)$ denotes the image of $T$ under the natural
action of $\mathfrak{S}_3$ on such $3$--tensors and $\varepsilon (\sigma)$ denotes the signature of $\sigma$.
\fi

\begin{remark}
Rewriting an instance of $\MCE$ as an instance of $\TFE$ or of $3$--tensor isomorphism shows that the problem is symmetric in the three parameters $m,n,k$. In particular, we may choose to permute $m,n,k$ as we like in our algorithm in order to minimize its complexity. Moreover, as we will explain in \Cref{lem:dualmce}, we may also switch $k$ for $mn-k$. 
\end{remark}

\subsection{Related works on attacks on $\MCE$ and $\ATFE$}\label{subsec:attack}

\subsubsection{About ALTEQ.}
In article \cite{ALTEQpaper} in which the design of ALTEQ is established, various cryptanalysis techniques are considered to solve $\ATFE$ problem. They include algebraic attacks: computing the matrix $P$ as the solution of a quadratic system; or MinRank based attacks (re discussed further as ``Leon--like techni\-ques''). The authors then consider a collision search attack with a cost $\widetilde{\OO}(q^{\frac{2n}{3}})$, which they claim to be the best possible. Finally,
NIST proposal ALTEQ \cite{ALTEQ} selects parameters with respect to 
a finer analysis of algebraic and MinRank based attacks.

\subsubsection{About MEDS.}
For the design of MEDS \cite{MEDSpaper}, the authors consider a graph-search based approach inspired from the works of Bouillaguet,
Fouque and V\'eber \cite{BFV13} on the polynomial isomorphism problem for spaces of quadratic forms.
This approach leads to an attack of complexity $\widetilde{\OO}(q^{\frac{2}{3}(m+n)})$. Also, they consider the possibility of algebraic modelling which turns out to be harder than
for $\ATFE$ since the unknown correspond to a pair of matrices $(P, Q)$
instead of a single one. They also study a ``Leon--like'' approach, 
a reference to Leon's algorithm \cite{L82} for determining code
equivalence that consists in harvesting minimum weight codewords
to determine the code equivalence. When transposed to the matrix code setting, the Hamming weight is replaced by the rank and such an approach
is nothing but the aforementioned MinRank based technique, which, following a recent result from Beullens \cite{Beullens21} on Hamming metric code equivalence, can be combined with a collision search technique.
MEDS' parameter selection rests on the complexity of both algebraic attacks
and Leon--like ones.

\subsubsection{Subsequent attacks.}
Recently, two attacks taking their inspiration from the Graph--based
techniques of Bouillaguet, Fouque and Véber \cite{BFV13} appeared 
in the literature.

\paragraph{Beullens' attack.}
First, Beullens proposed a graph--search--based technique to solve $\ATFE$. His attack turns out to be particularly efficient for small values of $n$. For instance, for $n$ odd, he could achieve a complexity in $\OO (q^{{(n-5)}/{2}}n^{11} + q^{n-7}n^6)$. This permitted to identify weak keys in \cite{ALTEQpaper}.

\paragraph{Narayanan, Qiao and Tang's attack.}
In \cite{nqt}, the authors introduced a new algorithm
solving both $\ATFE$ and $\MCE$ in the cubic case $k=m=n$. 
We conclude this section by sketching the principle of this algorithm
in order to point out the need for being in the cubic case.

As already explained in \S~\ref{subsec:related}, the problem
can be reformulated into that of the equivalence of two trilinear forms
\[
    f \colon \Fq^n\times \Fq^n \times \Fq^n \longrightarrow \Fq
    \quad \text{and}\quad
    g \colon \Fq^n \times \Fq^n \times \Fq^n \longrightarrow \Fq,
\]
where we look for a triple $P,Q,R \in \GL_n (\Fq)$
such that for any $x,y,z \in \Fq^n$, $g(x,y,z) = f(Px,Qy,Rz)$.

The idea of the algorithm consists first in guessing a pair $(x_1,x_1') \in {(\Fq^n)}^2$ such that $x_1' = Px_1$ and $f(x_1, \cdot, \cdot)$ is a bilinear form of rank $n-1$. Next, due to the rank constraint, from $x_1$ can be deduced a unique $y_1$ (up to scalar
multiplication) such that $f(x_1, y_1, \cdot) \equiv 0$.
Similarly, they deduce a $z_1$ such that $f(\cdot, y_1, z_1) \equiv 0$,
and an $x_2$ such that $f(x_2, \cdot, z_1) \equiv 0$ and so on.
By this manner, they construct 3 sequences $x_1, \dots, x_n$,
$y_1, \dots, y_n$ and $z_1, \dots, z_n$ for $f$ and similarly construct $x_1', \cdots, x_n'$,
$y_1', \dots, y_n'$ and $z_1', \dots, z_n'$ for $g$.
Stacking these vectors as columns of $n \times n$ matrices,
we get 6 matrices $X, Y, Z, X', Y'$ and $Z'$ that will be invertible
with a high probability.

The key observation is that 
\[X' = PX\quad Y' = QY \quad \text{and}\quad Z' = RZ.\]
Therefore, for any $x,y,z$
\[
    f(X'x, Y'y,Z'z) = f(PX x , QY y, QZz) = g(Xx,Yy,Zz).
\]
\alain{Verif pen and paper}
Thus, (up to some action of diagonal matrices that we do not discuss here) the trilinear forms $f_{x_1} \eqdef f(X'\ \cdot, Y'\ \cdot, Z'\ \cdot)$
and $g_{x_1'} \eqdef g(X\ \cdot, Y\ \cdot, Z\ \cdot)$ coincide.

In view of this observation, the algorithm solving $\MCE$
consists in a collision search between two dictionaries.
The first one collects pairs $(f_{x_1}, x_1)$, the left--hand term being the search key and the right--hand one being the corresponding value,
and the second one collects pairs $(g_{x_1'}, x_1')$. Once such a collision is found, determining
the equivalence becomes easy (see \cite{nqt} for further details). The space complexity of their algorithm is essentially the expected size of the computed dictionaries. Each dictionary has length $q^{(n-2)/2}$, and its entries are pairs consisting of a point of $\PP^n(\Fq)$ and a cubic 3-tensor of dimension $n\times n\times n$. Hence the total space complexity is $\OO(n^3q^{(n-2)/2})$.

\paragraph{Conclusion about Narayanan, Qiao and Tang.} It should be emphasized that the crux of their algorithm rests on the unique possibility (up to scalar multiplication) of passing from $x_i$ to $y_i$, from $y_i$ to $z_i$ and from $z_i$ to $x_{i+1}$. Such a technique is possible only because at
each step, the corresponding bilinear form is represented by a 
rank $n-1$ matrix of size $n \times n$. Hence, their approach strongly rests on the fact that they lie in the cubic case $k = m = n$.

\paragraph{Ran and Samardjiska's attack.} In \cite{ran24}, the authors describe a graph-based algorithm which solves the 3-tensor isomorphism problem in a specific case, namely when the graphs of the two isomorphic tensors contain cycles of length 3. This only happens in $\sim 1/q$ of all instances. However, contrary to \cite{nqt}, they are not limited to the cubic case. Their article contains a generic algorithm for the tensor isomorphism problem, and two versions adapted specifically to the $\MCE$ and $\ATFE$ problems. They all rely on the same \emph{modus operandi}: \begin{itemize}
    \item first, model by algebraic equations the existence of triangles in the graphs associated with the two tensors, and solve these using Gröbner basis techniques in order to construct lists of triangles found in each tensor's graph;
    \item then, for each pair of triangles in the two lists, try to construct the isometry by solving a system of linear and quadratic equations.
\end{itemize} 
The authors provide timings showing that their attack is more efficient with real-world parameters than the previous ones, even going as far as breaking the designed Level I parameters of ALTEQ in under half an hour. However, this only applies to those $1/q$ of all instances in which the considered tensors do have triangles in their associated graphs.

\paragraph{Comparison with the present work.} We do not expect our work to beat the time complexity of \cite{nqt} or \cite{ran24} in the cases where they apply. However, our algorithm solves the $\MCE$ problem in a very broad range of parameters, and is not restricted to the cubic case like \cite{nqt} or to those (very rare) tensors whose graphs contain triangles like \cite{ran24}. Moreover, it improves upon the space complexity of \cite{nqt} by a factor $\Theta(n^2)$.

\section{Technical overview}
\label{sec:techover}

In this article, we propose an algorithm to solve $\MCE_{m,n,k}(\CC,\DD)$ for the range of parameters $m,n,k$ such that $n\geqslant m$ and
\[
   k<m^2-1 \text{ or } mn-k<m^2-1
\]
(see \Cref{rk:constraintsk}). This includes the cubic case,
\emph{i.e.}, $k=m=n$, in which the complexity of our algorithm turns out to be similar
to that of \cite{nqt}.

\subsection{Preliminaries}
Our algorithm will use in a crucial way the notion of \emph{dual matrix
code} and that of \emph{Hull}. We give both definitions below. 
Recall that the \emph{trace} $\Tr(M)$ of a square matrix $M$ is the sum of its diagonal coefficients.

\begin{definition}\label{def:dual_code}
Let $\CC\subset\Fqmn$ be a linear code. The \emph{dual} of $\CC$ is the code \[\Cp\eqdef \{M\in\Fqmn\mid \forall C\in \CC, \Tr(M^\top C)=0\}.\]
\end{definition}

\begin{definition}\label{def:hull}
Let $\CC\subset\Fqmm$ be a matrix code. We will call \emph{hull} of $\CC$ the code \[h(C)=\{M\in \CC\mid\forall C\in \CC,\Tr(MC)=0\}.\] 
\end{definition}

\begin{remark} Beware that the hull is \textbf{not} the intersection of $\CC$ with its dual as defined in \Cref{def:dual_code}. It is the intersection with another orthogonal subspace, this time with respect to the bilinear form 
\if\arXiv2
\[(X,Y)\mapsto \Tr(XY).
\]
\else
\((X,Y)\mapsto \Tr(XY).
\)
\fi
The definition of the hull (\Cref{def:hull}) is the only place of the article where this nonstandard bilinear form is used. Besides, every dual or orthogonal complement which appears in the article is taken with respect to the usual inner product
\if\arXiv2
\[(X,Y)\mapsto \Tr(X^\top Y).\]
\else
\( (X,Y)\mapsto \Tr(X^\top Y).\)
\fi
\end{remark}

The subsequent lemmas yield two key observations for our algorithms:
\begin{enumerate}
    \item if two codes are equivalent, so are their duals;
    \item if two codes are conjugate, so are their hulls.
\end{enumerate}

\begin{lemma} \label{lem:dualmce}
Let $\CC,\DD\subset\Fqmn$ be two $\Fq$-vector spaces, and $P\in\GL_m(\Fq)$, $Q\in\GL_n(\Fq)$ be matrices such that $\DD=P\CC Q^{-1}$.
Then \[ \Dp=(P^{-1})^\top\Cp Q^\top.\]
\end{lemma}

\begin{proof}
Since $\CC$ and $\DD$ have the same dimension, so do $\Dp$ and $(P^{-1})^\top\Cp Q^\top $. Hence, it is enough to prove that one of these spaces is included in the other. Consider any $B\in \DD$ and $A\in \Cp$. There is a matrix $C\in\CC$ such that $B=PCQ^{-1}$. We have:
\begin{align*} \Tr(B^\top (P^{-1})^\top AQ^\top )&= \Tr((Q^{-1})^\top C^\top P^\top (P^{-1})^\top AQ^\top ) \\
&= \Tr((Q^{-1})^\top C^\top AQ^\top ) \\
&= \Tr(C^\top A) = 0. \tag{since $A\in \Cp$}
\end{align*}
Hence, \( \DD\subseteq (P^{-1})^\top \CC^\perp Q^\top.\)\qed
\end{proof}

\begin{lemma}\label{lem:hconj} Let $\CC,\DD\subset \Fqmm$ be two $\Fq$-vector spaces. Let $P\in\GLmq$ be a matrix such that $\DD=P\CC P^{-1}$. Then
\[ h(\DD) = Ph(\CC)P^{-1}.\]
\alain{pen and paper}
\end{lemma}

\begin{proof}
Let $C\in h(\CC)$, and set $D=PCP^{-1}\in\DD$. Let us show that $D\in h(\DD)$.
Let $B\in \DD$. There exists $A\in\CC$ such that $B=PAP^{-1}$. We have \begin{align*}
\Tr(BD)&=\Tr(PAP^{-1}PCP^{-1})   \\
&=\Tr(AC) =0. \tag{since $C\in h(\CC)$}
\end{align*}
The other inclusion is proved in the same way. \qed
\end{proof}

The following proposition, which says that roughly $1/q$ of all codes have a one-dimensional hull, is a consequence of results presented in \cite{dimhull}. It is explained in \Cref{sec:hulldim}, and will be crucial in the complexity analysis. 

In the sequel, we denote by $\ker (\Tr)$ the subspace of $\Fq^{m \times m}$
of matrices whose trace is zero.

\begin{restatable}{proposition}{hulldimprop}\label{hulldimprop} The proportion of $m \times m$ matrix codes contained in $\ker(\Tr)$ and whose hull has dimension 1 is asymptotically equal to \[ \frac{1}{q}\left( 1+\OO\left( \frac{m^2}{q^{(m^2-1)/2}}\right)\right).\] 
\end{restatable}

\subsection{Summary of the algorithm}

We are given two $k$-dimensional subspaces $\CC,\DD$ of $\Fqmn$. Our aim is to find two matrices $P\in\GL_m(\FF_q)$ and $Q\in\GL_n(\FF_q)$ verifying $\DD=P\CC Q^{-1}$. If we have found a suitable matrix $P$, computing $Q$ can be done using linear algebra (see \Cref{sec:recoverQ}). The strategy for finding $P$ consists first in guessing a pair
$(A, B) \in \CC^\perp \times \DD^\perp$ such that $B = (P^{-1})^\top AQ^{\top}$, that is, a pair $A, B$ which match with respect to the equivalence $\DD^\perp = (P^{-1})^\top C^\perp Q^\top$ given by \Cref{lem:dualmce}. With such a pair
at hand, one can reduce the equivalence problem to the conjugacy problem of the codes
\begin{equation}\label{eq:CA_CB}
    \CC_A \eqdef \CC A^\top \qquad \text{and} \qquad \DD_B \eqdef \DD B^\top.
\end{equation}
Indeed, if $B = PAQ^{-1}$ we prove in \Cref{lem:key_lemma} further that $\DD_B=P\CC_A P^{-1}$. Solving a matrix code conjugacy problem in this context is generally as hard as solving $\MCE$ \cite{grochow2023complexity}, but it is easy in a particular case: when the hull of both codes has dimension 1, we may easily find conjugate generators of these two hulls. The two main steps in order to find $P$ are the following.

\begin{enumerate}
    \item From $\CC, \DD$, construct two conjugate codes $\CC_A,\DD_B$ with one-dimensional hull. 
    \item Compute a matrix $R$ that conjugates these hulls and deduce a matrix $P$ such that that $\DD_B=P\CC_A P^{-1}$.
\end{enumerate}

\paragraph{First step.} We begin by 
finding two matrices $A\in\Cp$ and $B\in\Dp$ such that the codes
$\CC_A$ and $\DD_B$ of \eqref{eq:CA_CB} have conjugate hulls. For any $A$, one may find at least one such $B$, which is $(P^{-1})^\top AQ^\top $.

In order to determine these matrices $A$ and $B$, we construct a dictionary whose keys are (normalized and suitably chosen) polynomials $\chi\in\Fq[t]$ of degree $m$. The values corresponding to a key $\chi$ are the pairs $(A,U)\in\Fqmn\times\Fqmm$ such that the hull $h(\CC_A)$ is one-dimensional and generated by the matrix $U$ with characteristic polynomial $\chi$. 
Then, we apply the same process to $\DD$ and look for collisions.
This step is explained in detail in \Cref{sec:reduction}. 

\paragraph{Second step.} Once we have a pair of matrices $(A,B) \in \Cp \times \Dp$ such that $h(\CC_A)$ and $h(\CC_B)$ are one-dimensional and
generated by conjugate matrices $U$ and $V$, we may easily compute a
matrix $R\in\GLmq$ such that $V=RUR^{-1}$. We also impose in the collision search
that the characteristic polynomial $\chi$ of $U, V$ is squarefree so that $U, V$
are both diagonalizable in an extension of $\Fq$.
In this context, we will observe that the matrix $P$ we are looking for, \emph{i.e.} the one such that $\DD_B=P\CC_A P^{-1}$, is the product of $R$ by some
invertible matrix which can be expressed as $f(U)$ for some polynomial $f\in\Fq[t]$
 of degree less than $m$. The calculation of this polynomial is explained in \Cref{sec:find}.

\subsection{A comment on matrix code equivalence \emph{v.s.} matrix code conjugacy}
A remark that arises from our work, is that the equivalence problem
seems to become much easier when reducing from general matrix code equivalence
(\emph{i.e.}  arbitrary $P, Q$) to matrix code conjugacy (\emph{i.e.}
$m=n$ and $P=Q$). It is interesting to observe that from a complexity
theory point of view the two problems are polynomially equivalent
\cite[Thm.~A]{grochow2023complexity}. Still, the use of the hull gives a heuristic
polynomial-time algorithm that solves a proportion $\OO (1/q)$ of instances of the
conjugacy problem (the $1/q$ coming from the fact that a random matrix
code has a one-dimensional hull with probability $\OO (1/q)$).

This phenomenon could be compared with what happens in classical
coding theory, where two problems arise : the \emph{permutation
  equivalence problem} (finding a permutation matrix sending a code to
another) and the \emph{monomial equivalence problem} (finding a monomial
matrix, \emph{i.e.} the product of a permutation matrix and a nonsingular
diagonal matrix sending one code to another). When the ground
field cardinality $q$ is polynomial in the code length, the two
problems are known to be polynomially equivalent \cite{SS13} but
Sendrier's \emph{Support Splitting algorithm} \cite{S00} is on average efficient on
the former while being completely inefficient on the latter.

\subsection{Complexity and impact}

The complexity of the algorithm is dominated by the collision search in the first step. Given a uniformly random code $\CC$ and a uniformly random full-rank matrix $A\in\Cp$, the codes $\CC_A$ are uniformly distributed among the matrix codes in $\ker(\Tr)\subset\Fqmm$. Among these, a proportion of approximately $1/q$ have a one-dimensional hull. By an argument similar to the birthday paradox, computing two lists of length roughly $q^{(\kp-2)/2}$ is enough to find some collisions. This requires picking matrices $A$, and for each of these, computing the hull of $\CC_A$. The total time complexity, given in \Cref{th:complex}, is \[ \widetilde{\OO}\left( q^{\max(\frac{\kp}{2}, \kp-m+2)}\right) \]
operations in $\Fq$, where $\kp\eqdef mn-k$ is the dimension of the dual code $\CC^\perp$. In the cubic case $m=n=k$, by switching $\CC$ and $\Cp$, this complexity can be reduced to
\if\arXiv2
{\[ \widetilde{\OO}\left(q^{n/2}\right).\]
\else{\( \widetilde{\OO}\left(q^{n/2}\right).\)
\fi
The space complexity is \[ {\OO}\left((\kp+m+1)q^{\min(\frac{\kp}{2}-1, m-3)}\right)\]
elements of $\Fq$, which is essentially the size of the computed dictionaries. In the cubic case, this amounts to
\if\arXiv2{
\[ \OO\left(nq^{\frac{n}{2}-1}\right)\]
}
\else{
\( \OO\left(nq^{\frac{n}{2}-1}\right)\)
}
\fi
elements of $\Fq$. This reduces the space complexity of \cite{nqt} by a factor $n^2$, which, with the initial cubic parameters of MEDS-128, is about 200.

\section{Reducing to probably conjugate spaces}\label{sec:reduction}

We look at $k$-dimensional matrix codes inside $\Fqmn$. Given a code of dimension $k$, we will denote by $\kp=mn-k$ the dimension of its dual. 

\subsection{Structure of the reduction}

\Cref{lem:dualmce} shows that the instances $(\CC,\DD)$ and
$(\Cp,\Dp)$ of $\MCE$ are equivalent. In particular, for complexity
reasons, we may switch $(\CC,\DD)$ for $(\Cp,\Dp)$: we will
systematically choose the instance with the highest dimension. Indeed,
since collision search is performed on the dual codes, we fit in the
situation where the codes have the smallest possible duals. Thus from now on,
we suppose
\[
    \kp \leq k.
\]

\Cref{lem:dualmce} also shows that given $A\in \Cp$, the matrix $B=(P^{-1})^\top A Q^\top $ belongs to $\Dp$. The key of the algorithm lies in the following lemma.
\begin{lemma}\label{lem:key_lemma} Let $(A, B) \in \CC^\perp \times \DD^\perp$ such that
$B = (P^{-1})^\top A Q^\top$. Then, the codes
\if\arXiv2
\[\CC_A\eqdef \CC A^\top \quad \text{and} \quad \DD_B\eqdef \DD B^\top\]
satisfy
\[
  \DD_B = P\CC_A P^{-1}.
\]
\else
\[\CC_A\eqdef \CC A^\top \quad \text{and} \quad \DD_B\eqdef \DD B^\top
\quad
\text{satisfy}
\quad
  \DD_B = P\CC_A P^{-1}.
\]
\fi
\end{lemma}

\begin{proof}
This is a straightforward computation. \qed
\end{proof}

The aim of the first step of our algorithm is to find pairs
$(A,B)$ such that $\CC_A$ and $\DD_B$ are two conjugate
$k$-dimensional codes, in order to find $P$. Given any $\CC_A$,
$\DD_B$, finding a matrix $P$ such that $\DD_B=P\CC_A P^{-1}$ is
complicated: this is the code conjugacy problem (see for instance
\cite{grochow2023complexity}). However, it is much easier if one knows a
distinguished pair of conjugate elements $U\in\CC_A$, $V\in\DD_B$. In
order to find such a pair $(U,V)$, we need to find conjugate
one-dimensional subspaces in both $\CC_A$ and $\DD_B$. We can do this
when the hulls of both $\CC_A$ and $\DD_B$ are one-dimensional, since,
as shown in \Cref{lem:hconj}, the hulls of two conjugate matrix codes
are conjugate.

\begin{remark}\label{rk:constraintsk} By construction, there is an inclusion $\CC_A \subset \ker (\Tr)$. Moreover, in order to have a one-dimensional hull, we need this inclusion to be a strict. Since we require $\CC_A$ to have the same dimension as $\CC$, this means that our algorithm in this form only works for
  \[k<m^2-1.\]
\end{remark}

Given an instance $(\CC,\DD)$ of $\MCE_{m,n,k}$, our goal
is to find matrices $P\in\GLmq$ and 
$Q\in\GLnq$ such that $\DD =P\CC Q^{-1}$.  The method below allows us to reduce the problem to the case where $m=n$, $P=Q$, and $\CC$ and $\DD$ have non trivial conjugate hulls.
The reduction consists in the following steps. \begin{enumerate}
    \item Construct a dictionary $\{ \chi : (A,U)\}$, where
    $\CC_A = \CC A^\top$ has a one--dimensional hull,
    $U\in\Fqmm \setminus \{ 0\}$ generates this hull and $\chi\in\Fq[t]_{\leq m}$ is the characteristic polynomial of $U$, which we require to be squarefree. This is done in a very straightforward way: pick $A$ at random, compute $h(\CC_A)$ and if it is one-dimensional and generated by a matrix $U$, compute its characteristic polynomial $\chi_U$ and add the entry $(\chi\colon (A,U))$ to the dictionary. To make the second step easier, we only keep $A$ when $\chi$ is separable. The precise procedure is explained in Algorithm~\ref{alg:constrdict}.  In order to find collisions more easily, we normalize the characteristic polynomials as explained in \Cref{sec:normalize}: this reduces the number of possible characteristic polynomials to approximately $q^{m-3}$ (see \Cref{lem:nbpoints}).
    \item Pick random matrices $B\in\DD^\perp$, and if the hull of $\DD B^\top $ is one-dimensional, check if the characteristic polynomial of one of its generators is a key in the dictionary.  The aforementioned conditions on the characteristic polynomials directly imply that the generators $U,V$ of $h(\CC_A),h(\DD_B)$ having the same characteristic polynomial $\chi$ are conjugate: for each collision, we immediately compute $R\in\GLmq$ such that $V=RUR^{-1}$. The collision-finding procedure is described in Algorithm \ref{alg:findcoll}.
\end{enumerate}
This yields a list of tuples $(A,B,U,V,R)$ such that the codes $\CC_A$ and $\DD_B$ have one-dimensional hulls respectively generated by matrices $U,V$ such that $V=RUR^{-1}$. To these tuples, we then apply an algorithm of \Cref{sec:find} which allows to find a suitable matrix $P$ such that $\DD_B = P \CC_A P^{-1}$. In order to compute $Q$, we now need to solve $\DD=(P\CC)\cdot Q$, where $P\CC$ is known. This is an easy problem, which is solved by linear algebra as explained in \Cref{sec:recoverQ}.
\begin{figure}[h]
\begin{algorithm}[H]\label{alg:charpoly}
\SetAlgoLined  
\caption{\textsc{ComputeNormalizedCharpoly}}
\KwData{$k$-dimensional code $\CC\subset\ker(\Tr)\subset\Fqmm$ such that $\dim h(\CC)=1$}
\KwResult{Pair $(\chi,U)$ where $U \in\Fqmm$ generates $h(\CC)$, and $\chi\in\Fq^{m-2}-\{ 0\}$ represents $U$'s characteristic polynomial}
\hrulefill \\
Compute a generator $U$ of $h(\CC)$\\
Compute char. polynomial $a_0+a_1t+\dots +a_{m-3}t^{m-3}+t^m$ of $U$\\
Set $\chi=(a_{m-3},a_{m-4},\dots,a_0)\in\Fq^{m-2}-\{ 0\}$\\
$\lambda=$\textsc{Normalize}$(U,\chi)$ using Algorithm \ref{alg:normalize}\\
\Return $(\lambda\diamond \chi,\lambda U)$ \hfill (where $\diamond$ is defined further below Eq.~($\star\star\star)$)
\end{algorithm}
\caption{computes the normalized generator of the hull of a code with one-dimensional hull.}
\end{figure}

\begin{figure}[!h]
\begin{algorithm}[H]\label{alg:constrdict}
\SetAlgoLined  
\caption{\textsc{ConstructDict}}
\KwData{$k$-dimensional code $\CC\subset\Fqmn$, integer $L$}
\KwResult{Dictionary $\{ \chi:(A,U)\}$ with $L$ keys\\ where $U \in\Fqmm$ generates $h(\CC_A)$ and has characteristic polynomial $\chi$}
\hrulefill \\
$\Dict=\{ \}$\\
Compute a basis of $\CC^\perp$\\
\While{$length({\rm Dict})<L$}{
	Pick a random $A\in\CC^\perp$\\
	\If{${\rm rk}(A)=m$ and $\dim(\CC_A)=k$}{
	    \If{$\dim h(\CC_A)=1$}{
	        $(\chi,U)={\rm ComputeNormalizedCharpoly}(\CC_A)$\\
	        \If{${\rm gcd}(\chi(t),\chi'(t))=1$ \textbf{and} $\chi \notin \Dict$}{
              Add entry $(\chi:(A,U))$ to $\Dict$
	    }}}}
\Return $\Dict$
\end{algorithm}
\caption{constructs a dictionary whose keys are separable polynomials, and whose values are pairs of matrices $(A,U)$ such that $h(\CC_A)=\Fq U$.}
\end{figure}

\begin{figure}[H]
\begin{algorithm}[H]\label{alg:findcoll}
\SetAlgoLined  
\caption{\textsc{Finding $P$}}
\KwData{$k$-dimensional code $\DD\subset\Fqmn$, integer $N$, dictionary $\rm Dict$}
\KwResult{Triple $(A,B,P)\in(\Fqmn)^2\times(\Fqmm)^2\times\GLmq$ s.t. $\mathcal{D}_B = P \mathcal{C}_A P^{-1}$}
\hrulefill \\
$i = 0$\\
\While{$i < N$}{
	Pick a random $B\in\DD^\perp$\\
	\If{${\rm rk}(B)=m$ \textbf{and} $\dim(\DD_B)=k$}{
\If{$\dim h(\DD_B)=1$}{
        $i = i+1$\\
	        $(\chi,V)={\rm ComputeNormalizedCharpoly}(\DD_B)$\\
	        \If{$\chi$ is a key of $\rm Dict$ with value $A$}{     
Use one of the Algorithms of Section~\ref{sec:find} to deduce $P$\\
            \If{success}
            {\Return $(A,B,P)$}
          }}}} \smallskip
    \Return $\bot$
\end{algorithm}
\caption{returns a triple $(A,B,P)$ where $\DD_B=P \CC_A P^{-1}$.}
\end{figure}

\subsection{Distribution of the computed matrix spaces and polynomials}
\label{sec:distrib}
 In this section, we discuss the distribution of the matrix spaces and characteristic polynomials obtained using the algorithms above. Given a vector space $V$ and an integer $d$, we denote by $\Gr_d(V)$ (resp. $\Gr_{\leq d}(V))$ the set of all $d$-dimensional (resp. at most $d$-dimensional) linear subspaces of $V$. We prove that given a uniformly random $\CC\in\Gr_k(\Fqmn)$ and $A\in\Cp$ such that $\CC_A$ has one-dimensional hull (and some mild additional conditions), the distribution of the characteristic polynomials of a generator of these hulls is asymptotically uniform with respect to $q$.\\

 Given a matrix $A\in\Fqmn$, we define the map
 \begin{align*} \phi_A\colon \Fqmn&\longrightarrow\Fqmm \\ M&\longmapsto MA^\top.   \end {align*}
For a $k$-dimensional code $\CC\subset\Fqmn$ and a matrix $A\in\Fqmn$, we consider in our reduction the code $\phi_A(\CC)=\CC_A\subset\Fqmm$. This amounts to considering the map 
\begin{align*} \Phi\colon \Gr_k(\Fqmn)\times\Fqmn&\longrightarrow \Gr_{\leq k}(\Fqmm) \\
(\CC,A)&\longmapsto \phi_A(\CC)=\CC_A.
\end{align*}
We will choose $A$ to have full rank $m$ (recall that $m\leq n$). This entails that $\phi_A$ is surjective. The preimages of a $k$-dimensional code $\DD\subset\Fqmm$ under $\Phi$ are exactly the pairs $(\CC,A)$ such that $\CC\subset \phi_A^{-1}(\DD)$ and $\CC\cap\ker(\phi_A)=0$. We now set
\[X\eqdef\{ (\CC,A)\in \Gr_k(\Fqmn)\times\Fqmn\mid A\in\CC^\perp,\,\mathrm{rk}(A)=m,\,\CC\cap\ker(\phi_A)=0\}.\]

\begin{lemma}\label{lem:f1eq} The restricted map \[f_1=\Phi_{|X}\colon X\to \Gr_k(\ker(\Tr))\tag{$\star$},\]
where $\Gr_k(\ker(\Tr))$ denotes the set of $k$--dimensional spaces of $m \times m$
matrices whose trace is zero,
is surjective and equidistributed (i.e. each element of its image has the same number of preimages).
\end{lemma}

\begin{proof}
Given $\DD\in\Gr_k(\Fqmm)$ whose elements have trace zero, any element $(\CC,A)\in\Phi_{|X}^{-1}(\DD)$ satisfies $A\in\CC^\perp$. Given a rank $m$ matrix $A\in\Fqmn$, the codes $\CC$ such that $\CC A^\top=\DD$ and $(\CC,A)\in X$ are exactly the complementary subspaces of $\ker(\phi_A)$ in $\phi_A^{-1}(\DD)$.
The number of elements in $\Phi_{|X}^{-1}(\DD)$ is the number of rank $m$ matrices in $\Fqmn$ multiplied by the number of complementary subspaces of an $m(n-m)$-dimensional subspace in a $(k+m(n-m))$-dimensional $\Fq$-vector space. The latter number is nonzero and does not depend on a particular choice of $\DD$. Hence, the map $f_1$ is surjective and equidistributed.
\qed\end{proof}

\begin{lemma}
In a code $\CC_A\subset\ker(\Tr)$, any element in the hull of $\CC_A$ satisfies $\Tr(U^2)=0$.
\end{lemma}

\begin{proof}
This is a direct consequence of the definition of the hull (Definition~\ref{def:hull}).
\end{proof}

\begin{lemma}\label{lem:f2eq} If $q$ is not a power of 2, the map \[ f_2\colon \{ \CC\in\Gr_k(\ker(\Tr)) \mid \dim h(\CC)=1\} \to \{ U\in\Fqmm\mid \Tr(U)=\Tr(U^2)=0\}/\Fq^\times  \tag{$\star\star$}\]
which sends $\CC$ to a generator of $h(\CC)$ (modulo the action of $\Fq^{\times}$) is equidistributed. In particular, as soon as the set on the left is nonempty, $f_2$ is surjective.
\end{lemma}

\begin{proof}  We work in $\ker(\Tr)$ with the non-degenerate bilinear form $(X,Y)\mapsto \Tr(XY)$.
Consider any two matrices $U_1,U_2$ such that $\Tr(U_i)=\Tr(U_i^2)=0$. To prove equidistribution, it is enough to construct a bijection between their preimages under this map. The map $\Fq U_1\to \Fq U_2$ which sends $U_1$ to $U_2$ is an isometry with respect to the aforementioned bilinear form. Since ${\rm char}(\Fq)\neq 2$, Witt's extension theorem \cite[Thm.~5.2]{groups} ensures that this map extends to an isometry $g$ of $\ker(\Tr)$. Then, the map\[ \{ \CC\subset\ker(\Tr)\mid h(\CC)=\Fq U_1\} \to \{ \CC\subset\ker(\Tr)\mid h(\CC)=\Fq U_2\} \] which sends $\CC$ to $g(\CC)$ is a bijection.
\qed\end{proof}

\begin{remark}
Recall that \Cref{hulldimprop} states that asymptotically, $1/q$ of all matrix codes in $\Gr_k(\ker \Tr)$ have a one-dimensional hull. Hence, the number of these codes is equivalent to $q^{k(m^2-1-k)-1}$ when $q\rightarrow \infty$.
Therefore, for big enough $q$, such codes exist, and the map $f_2$ is always surjective. 
\end{remark}

\begin{lemma}\label{lem:repartcharpoly} Let $\chi\in\Fq[t]$ be a separable polynomial of degree $m$ such that $\chi(0)\neq 0$. The number of matrices $U$ with characteristic polynomial $\chi$ is asymptotically (when $q\to\infty$) equivalent to $q^{m^2-m}$.
\end{lemma}

\begin{proof} Let $U$ be a matrix with characteristic polynomial $\chi$. Since $\chi$ is separable, it is also the minimal polynomial of $U$, and the matrices with characteristic polynomial $\chi$ are conjugates of $U$. There are as many conjugates of $U$ as elements in the quotient
  \[\GL_m(\FF_q)/\{ P\in\GL_m(\Fq)\mid PUP^{-1}=U\}.\] Since $U$ has a separable characteristic polynomial, any matrix which commutes with $U$ is a polynomial in $U$ \cite[Cor.~IV.E.8]{kerr}. We are looking for the cardinality of $\Fq[U]\cap \GL_m(\Fq)$. 
  A classical consequence of Cayley Hamilton theorem entails that $\Fq[U]\cap \GL_m(\Fq)$
  is nothing but the group $\Fq [U]^\times$ of invertible elements of the ring $\Fq[U]$.
  Hence, the polynomials $f\in\Fq[t]/(\chi)$ such that $f(U)$ is not invertible are those that are divisible by an irreducible factor of $\chi$. Their number is maximal when $U$ is diagonalizable over $\Fq$, in which case there are less than $mq^{m-1}$ such polynomials. Hence, $\Fq[U]\cap \GL_m(\Fq)$ has at least $q^m-mq^{m-1}=q^m(1-m/q)$ elements; since it always has less than $q^m$ elements, its cardinality is equivalent to $q^m$, and that of $\GL_m(\Fq)/(\Fq[U]\cap\GL_m(\Fq))$ is equivalent to $q^{m^2-m}$.
\qed\end{proof}

Consider the set of matrices $U\in\GLmq$ such that $\Tr(U)=\Tr(U^2)=0$, up to scalar multiplication. The characteristic polynomial $\chi_U\in\Fq[t]$ of such a matrix $U$ is of the form $t^m+a_{m-3}t^{m-3}+\dots+a_1t+a_0$, with $a_0\neq 0$ since $U$ is invertible.  For any $\lambda\in\Fq^\times$, the characteristic polynomial of $\lambda U$ is \[ \chi_{\lambda U}=t^m+\lambda^3 a_{m-3}+\dots+\lambda^{m-1}a_1+\lambda^m a_0.\] 
Hence, there is a  map
\begin{align*}
f_3\colon \{ U\in \GLmq\mid \Tr(U)=\Tr(U^2)=0\}/\Fq^\times &\longrightarrow \Fq^{m-2}/\Fq^\times\\
U&\longmapsto (a_{m-3},\dots,a_0) \tag{$\star\star\star$}
\end{align*}
where $\Fq^\times$ acts on $\Fq^{m-2}$ via $\lambda\diamond (a_{m-3},\dots,a_0)=(\lambda^3a_{m-3},\dots,\lambda^m a_0)$. Its image is $(\Fq^{m-2}-(\Fq^{m-3}\times\{ 0\}))/\Fq^\times$.
\Cref{lem:repartcharpoly} asserts that any element of the form $(a_{m-3},\dots ,u_0)$ corresponding to a separable polynomial has $\sim q^{m^2-m}$ preimages under $f_3$. \\

Denote by \[
  \Sep\subset (\Fq^{m-2}-(\Fq^{m-3}\times\{ 0\}))/\Fq^\times\] the set of classes of separable characteristic polynomials with nonzero constant coefficient.

\begin{remark} We give more details about this construction in \Cref{sec:normalize}. In particular, we show in \Cref{lem:nbpoints} that the set $\Sep$ has $\sim q^{m-3}$ elements. In Algorithm \ref{alg:charpoly}, we use a unique representative of each class of characteristic polynomials. The way of computing such a normalized representative is also explained in \Cref{sec:normalize} and presented in Algorithm \ref{alg:normalize}.
\end{remark}

\begin{proposition}\label{prop:equidis}  Suppose $q$ is not a power of $2$.
Denote by \[f_q=f_3\circ f_2\circ f_1\colon X\to \Fq^{m-2}/\Fq^\times\] the map which sends $(\CC,A)$ to the equivalence class of the tuple of coefficients of the characteristic polynomial of a generator of the hull of $\CC_A$. The maps $f_1,f_2,f_3$ are defined in $(\star),(\star\star),(\star\star\star)$.
The map \[ {f_q}_{|f_q^{-1}(\Sep)}\colon f_q^{-1}(\Sep)\to \Sep\]
is asymptotically equidistributed, i.e.
\[ \min_{\chi\in\Sep} |f_q^{-1}(\chi)| \underset{q\to\infty}{\sim} \max_{\chi\in\Sep}|f_q^{-1}(\chi)|.\]
\end{proposition}

  \begin{proof}  It is a direct consequence of Lemmas~\ref{lem:f1eq}, \ref{lem:f2eq}
  and \ref{lem:repartcharpoly}.
\qed\end{proof}

\begin{remark}\label{rem:equidis} The elements produced by Algorithm \ref{alg:constrdict} are characteristic polynomials obtained by picking uniformly random elements of $f_q^{-1}(\Sep)$ and computing a normalized representative of their image under the map $f$.
Hence, \Cref{prop:equidis} shows that given uniformly random inputs $\CC\in\Gr_k(\Fqmn)$, the distribution of normalized characteristic polynomials $\chi\in \Sep$ produced by Algorithm \ref{alg:constrdict} is asymptotically uniform. 
\end{remark}

\begin{remark}\label{rem:equisqare} The result above shows that the distribution of the computed characteristic polynomials is asymptotically uniform for random $\CC$ and $A$. But in practice, a fixed code $\CC$ is given to us. In that case, we have not said anything about the distribution of the codes $\CC A^\top $ yet. 
In the special case $m=n$, the map \[\psi_\CC\colon \Cp\cap\GL_m(\Fq)\to\Gr_k(\Fqmm)\] sending $A$ to $\CC A^\top $ is equidistributed. 
Indeed, given two codes $\DD_1=\CC A_1^\top \in\Gr_k(\Fqmm)$ and $\DD_2=\CC A_2^\top $,
the map
  \begin{align*}
    \psi_\CC^{-1}(\DD_1)& \longrightarrow \ \psi_\CC^{-1}(\DD_2) \\
    B_1 &\longmapsto \ A_2 A_1^{-1}B_1
  \end{align*}
is a bijection. 
\alain{pen and paper}
\end{remark}

\subsection{Complexity analysis}\label{sec:complexity}

In this section, the symbol $\sim$ always denotes asymptotic equivalence, and the notation $o(\cdot)$ denotes asymptotic domination, with respect to the parameter $q$. 
We make the following assumption, justified by Remarks \ref{rem:equidis} and \ref{rem:equisqare}.
\begin{assumption}\label{assum} Given a code $\CC$, distinct full-rank matrices $A$ yield distinct codes $\CC_A=\CC A^\top$ and the characteristic polynomials $\chi$ are uniformly distributed among the codes $\CC_A$ with one-dimensional hull. For this, we require $k\leq m^2-2$: otherwise, the codes $\CC A^\top $ would be the full $\ker(\Tr)$ as soon as $A$ has full rank, and could not have a one-dimensional hull.
\end{assumption} We are going to answer the following questions: \begin{enumerate}
\item How many matrices $A$ do we need to sample to find enough characteristic polynomials?
    \item How many operations are needed to compute the dictionary?
    \item What is the total complexity of running Algorithms \ref{alg:constrdict} and \ref{alg:findcoll} with the parameters answering the previous questions?
\end{enumerate}

\subsubsection{How many matrices $A$ do we need to sample in order to find enough characteristic polynomials?} 

For any $A\in\Cp$ and any $\lambda\in \Fq^\times$,
$\CC_A=\CC_{\lambda A}$. Hence, the total number of codes $\CC_A$,
$A\in \Cp$ is less than
\[ \frac{1}{q-1}(\#\Cp-1)=\frac{q^\kp-1}{q-1} \sim q^{\kp-1}\] where
$\kp=\dim(\Cp)=mn-k$. The number of $\CC_A$ with one-dimensional hull
is therefore equivalent to $q^{\kp-2}$ by \Cref{hulldimprop}. The
total number of possible classes of separable characteristic
polynomials is $\sim q^{m-3}$ (see \Cref{lem:nbpoints}).

Thus, the dictionary constructed in Algorithm~\ref{alg:constrdict} will
have size $L \leq q^{m-3}$. Then, according to the usual list collision arguments,
the number $L'$ of one-dimensional hulls to check in Algorithm~\ref{alg:findcoll} should satisfy
\[LL' \sim q^{\kp - 2}.\]

We have to treat two cases separately:
\begin{enumerate}[$(i)$]
\item\label{case:1} $\kp - 2 \leq 2(m-3)$, where the dictionary constructed by Algorithm~\ref{alg:constrdict}
  will not need to cover all the possible characteristic polynomials and we take
  \[
    L \sim q^{\frac{\kp}{2}-1} \quad {\text{and}} \quad  L' \sim q^{\frac{\kp}{2}-1};
  \]
\item\label{case:2} $\kp - 2 > 2(m-3)$, where there are not enough different characteristic polynomials to have $L\sim L'$, and we take
  \[
    L \sim q^{m-3} \quad \text{and}\quad L' \sim q^{\kp-m+1}.
  \]
\end{enumerate}

\begin{lemma}\label{lem:nbmat} Let $r$ be an integer. The average number of matrices to sample in Algorithm \ref{alg:constrdict}  in order to get $r$ distinct characteristic polynomials
is $\sim qr$ if $r = o (q^{m-3})$ and $\sim qr\log(r)$ if $r\sim q^{m-3}$.
\end{lemma}
\begin{proof}
The latter case is a classical result usually called \emph{coupon collector's problem}, while for the former we prove a variant. Denote by \[N_\chi\sim q^{m-3}\] the total number of possible
  characteristic polynomials with the shape $X^m + a_{m-3}X^{m-3}+ \cdots + a_0$, by $M \leq (q^{\kp}-1)/(q-1)$ the number of
  elements in $(\Cp-\{ 0\})/\Fq^\times$ with full rank and by $S_r$ the number of
  matrices $A$ we have to sample in order to get $r$ different
  characteristic polynomials of matrices spanning one-dimensional
  hulls of codes $\CC_A$.  Denote by $s_j$ the number of matrices to
  sample after having a list of $j-1$ distinct polynomials in order to
  get the $j$-th one. We seek to compute the expected value
  \[\EE(S_r)=\EE(s_1)+\dots+\EE(s_r).\] The random variable $s_j$
  follows a geometric distribution: it is the first success of a
  Bernoulli variable. The parameter $p_j$ of this variable is computed
  as follows: it is the proportion, among all the elements of
  $(\CC^\perp-\{ 0\})/\Fq^\times$, of those (equivalence classes of)
  matrices $A$ yielding a code $\CC_A$ with one-dimensional hull and
  characteristic polynomial that is not among the $j$ polynomials
  already in the list. From \Cref{hulldimprop}, the number of full--rank matrices $A$ that yield a code
  $\CC_A$ with one-dimensional hull is
  \[M\left(\frac{1}{q}+\OO\left(
        \frac{m^2}{q^{(m^2+1)/2}}\right)\right).\] 
        Under Assumption~\ref{assum}, the number of
  matrices that yield a code $\CC_A$ with a one-dimensional hull and
  one of the $j$ characteristic polynomials already in the list
  is
  \[j\cdot \frac{M\left( \frac{1}{q}+\OO\left(
          \frac{m^2}{q^{(m^2+1)/2}}\right) \right)}{N_\chi}\cdot\] Hence
\begin{align*}
    p_j&=\frac{1}{M}\left[\frac{M}{q}+\OO\left( \frac{m^2M}{q^{(m^2+1)/2}}\right)-j\frac{M}{qN_\chi}+\frac{j}{N_\chi}\OO\left(\frac{m^2M}{q^{(m^2+1)/2}}\right)\right] \\
    &= \frac{1}{q}\left(1-\frac{j}{N_\chi} \right)+\OO\left( \frac{m^2}{q^{(m^2+1)/2}}\right)\\
    &\sim \frac{1}{q}\left( 1-\frac{j}{N_\chi} \right) \tag{since $\frac{1}{qN_\chi}\sim \frac{1}{q^{m-2}}$}\\
    &\sim \frac{N_\chi-j}{qN_\chi}.
\end{align*}
The expected value of the geometric random variable $s_j$ with parameter $p_j$ is $1/p_j$. Hence, using the fact that $r=o(N_\chi)$, \begin{align*}
\EE(S_r)&\sim qN_\chi\left( \frac{1}{N_\chi} +\dots+\frac{1}{N_\chi -r+1}\right) \\
&\sim qN_\chi \log\left( \frac{N_\chi}{N_\chi-r}\right)\\
&\sim -qN_\chi \log\left(1- r/N_\chi\right)
\sim qr.
\end{align*}
\qed\end{proof}

\subsubsection{Complexity of computing the dictionary}

\begin{lemma}\label{lem:compl2} The average complexity of Algorithm~\ref{alg:constrdict} with input a $k$-dimensional code $\CC\subset \Fqmn$ and a desired list length $L$ is \begin{align*}
    \OO(qLk(nm^{\omega-1}+k m^2)) \quad &\text{if}\ \ L=o(q^{m-3})\\
    \OO(q^{m-2}km(nm^{\omega-1}+k m^2)) \quad &\text{if}\ L \sim q^{m-3}.
  \end{align*}
\end{lemma}

\begin{proof} In order to get $L$ distinct characteristic polynomials,
  \Cref{lem:nbmat} tells us that we need to sample $\sim qL$ matrices
  $A$ if $L = o (q^{m-3})$ and $\sim qL\log L$ if $L \sim q^{m-3}$. In
  this last case, this requires to sample $\sim mq^{m-2}$
  matrices. For each of these, we first need to compute a basis
  $(C_1,\dots, C_k)$ of $\CC A^\top $, which is given by $k$ products
  of a matrix of size $m\times n$ by a matrix of size $n\times m$;
  this requires $\OO(knm^{\omega-1})$ operations in $\Fq$. Then, we
  need to compute $\CC_A\cap \CC_A^\perp$, which is given by the
  kernel of the (symmetric) Gram matrix
  $(\Tr(C_iC_j))_{1\leqslant i,j\leqslant k}$. Computing the diagonal
  entries of a given product $C_iC_j$ requires $\OO(m^2)$ operations
  in $\Fq$. Hence, the Gram matrix is computed in $\OO(k^2m^2)$
  operations in $\Fq$. Computing its kernel takes $\OO(k^\omega)$
  operations in $\Fq$.  When the hull has dimension 1, we then only
  need to compute the characteristic polynomial of the generator we
  have found, which is done in $\OO(m^{\omega})$ operations
  \cite[Thm.~1.1]{neiger}, and to normalize it. This normalization can
  be precomputed for a proportion $(1-2/q)$ of all cases (see
  \Cref{rk:precompute}).  So sampling one matrix takes
  $\OO(knm^{\omega-1}+k^2m^2+k^\omega+m^\omega)$ operations, which,
  since $k\leqslant m^2$ and $m\leqslant n$ gives
  $\OO (k(nm^{\omega-1}+km^2))$.  Multiplying this by the number of
  sampled matrices (mentioned in the beginning of the present proof)
  yields the result.\qed\end{proof}

\subsubsection{Total complexity of this reduction step}

Recall that we allowed ourselves to replace $\CC$ with $\Cp$ if needed, in order to ensure that $\kp\leq k$ and that we need $k<m^2-1$ for the algorithm to work (see \Cref{rk:constraintsk}).

\begin{theorem}\label{th:complex} Suppose that $m\leq n$ and $\kp\leqslant k<m^2-1$ and that Assumption~\ref{assum} holds. Under Assumption~\ref{ass:system},
Algorithm \ref{alg:findcoll}  takes an expected
  complexity of
\[
    \OO\left(km( nm^{\omega-1}+km^2) q^{\max(\frac{\kp}{2},\kp-m+2)} + m^{2\omega}q^{\kp -m+1}\right) 
\]  
  operations in $\Fq$, and a space complexity of \[ \OO\left((\kp+m+1)q^{\min (\frac{\kp}{2}-1, m-3)}\right)\] elements of $\Fq$.
\end{theorem}

\begin{remark}
 One can get rid of Assumption~\ref{ass:system} by replacing the techniques of Subsection~\ref{ss:poly_solve} by those of Subsection~\ref{ss:diagonalisation}
 at the cost of another sub-exponential term in the time complexity (Lemma~\ref{lem:comp_findP}). Namely, this would give an overall complexity of
 \[ \OO\left(km( nm^{\omega-1}+km^2) q^{\max(\frac{\kp}{2},\kp-m+2)} + q^{\kp -m+1}k^{\omega-1}m^{2}q^{3\sqrt{m}/\sqrt{\log m}\log q}\right)\]
 operations in $\Fq$.
\end{remark}

\begin{proof}
  We consider separately Cases~(\ref{case:1}) and (\ref{case:2})
  introduced in Page~\pageref{case:1}. The time complexity of
  Algorithm~\ref{alg:constrdict}, given by Lemma~\ref{lem:compl2}, is 
  \[\OO \left(k (nm^{\omega-1}+km^2) q^{\max(\frac{\kp}{2}, \kp-m+2)}\right).\]
  The cost of Algorithm~\ref{alg:findcoll} depends on the number of ``false positives'' encountered, \emph{i.e.} situations where we find for $h(\DD_B)$
  a characteristic polynomial which is an entry of the dictionary but running an Algorithm of Section~\ref{sec:find} shows that $\DD_B$ is not a conjugate of the code $\CC_A$ corresponding to that entry. For each  key of the dictionary, there are $\sim q^{\kp-2} / q^{m-3} = q^{\kp-m+1}$ one-dimensional hulls corresponding to this key, roughly all of which are false positives.

  In Case~(\ref{case:1}), Algorithm~\ref{alg:findcoll} 
  costs
  \[\OO  \left(km( nm^{\omega-1}+km^2) q^{\frac{\kp}{2}} + m^{2\omega}q^{\kp -m+1} \right).\]
  
  In Case~(\ref{case:2}), recall that the algorithms of Section~\ref{sec:find}, whose complexities are given in
  \Cref{lem:sys_calc_f}, have to be called on average $q^{\kp-m+1}$ times. Here, the complexity of calling the algorithms of Section~\ref{sec:find} outweighs that of computing the dictionary, and the time complexity of
  Algorithm~\ref{alg:findcoll} is
  \[
    \OO(m^{2\omega}q^{\kp-m+1} ).
  \]
  The space complexity is simple to compute: it is dominated by the number $L$ of entries in the dictionary. There are
  $\OO(q^{\min(\frac{\kp}{2}-1, m-3)})$ such entries, which consist of a matrix $A\in\Cp$ and a characteristic polynomial. To reduce space complexity, one
  may store in each of these entries only the coordinates of the
  matrix $A$ in a basis of $\Cp$ as well as the characteristic
  polynomial of the generator of $h(\CC_A)$, which amounts to
  $\kp+m+1$ field elements per entry.  \qed\end{proof}

\section{Finding the right matrix}
\label{sec:find}
In order to shorten the notations, we now denote by $\CC,\DD$ the codes $\CC_A,\DD_B$.
We are in the following situation: we are given two codes $\CC,\DD\subset\Fqmm$ with one-dimensional hulls generated respectively by matrices $U,V$ and a matrix $R\in\GLmq$ such that $V=RUR^{-1}$. Our aim is to decide whether $\CC_A$ and $\DD_B$ are conjugate by some matrix $P\in\GLmq$, and if they are, find such a matrix.

In this section, we propose two approaches. The first one is based on multivariate
polynomial system solving and has a polynomial time complexity conditioned by some assumption. The second one is based on diagonalization arguments, has a sub-exponential time complexity but does not rest on any assumption.

\subsection{First approach: using polynomial system solving}\label{ss:poly_solve}
The one-dimensional hulls of the codes $\CC$ and $\DD$ are respectively generated by conjugate matrices $U,V\in\GLmq$ having a squarefree characteristic polynomial. We can compute a matrix $R\in\GLmq$ such that $V=RUR^{-1}$, hence $h(\DD)=Rh(\CC)R^{-1}$. We are looking for a matrix $P\in\GLmq$ such that $\DD=P\CC P^{-1}$. 
We know that the matrix $P\in\Fqmm$ we are looking for
satisfies $V=PUP^{-1}$. Therefore, there exists a matrix $T\in\Fqmm$ which commutes with $U$ such that $P=RT$. Since the characteristic polynomial of $U$ is separable, we can write $T=f(U)$, where
$f=\alpha_0+\alpha_1t+\dots+\alpha_{m-1}t^{m-1}\in\Fq[t]$ (see
\cite[Cor.~IV.E.8]{kerr}). We know $R$ and search for a polynomial
$f$ such that \begin{equation} \DD=Rf(U)\CC f(U)^{-1}R^{-1}.\tag{$\circ$}\end{equation} 
A usual argument (for instance by Cayley-Hamilton Theorem) shows that there exists a polynomial $g$
of degree $<m$ such that $f(U)^{-1} = g(U)$. Then, we can solve the following system whose unknowns are the
coefficients of $f,g$:
\begin{equation}\label{eq:linearized}
Rf(U)\CC g(U)R^{-1} \subset \DD.
\end{equation}
The above system is bilinear in the $m$ coefficients of $f$ and the $m$ coefficients of $g$. Thus, when linearizing,
this yields $m^2$ unknowns while $\CC, \DD \in \Fq^{m \times m}$ both have dimension $k$ (recall that for short we denote
$\CC_A, \DD_B$ by $\CC, \DD$), hence the system has $k(m^2-k)$ equations which exceeds $m^2$ as soon as $1<k<m^2-1$.
Thus, the linearized system is over-constrained, which encourages to make the following assumption.

\begin{assumption}\label{ass:system}
The linearized version of System~\eqref{eq:linearized} has a space of solutions of dimension $1$ when  $\CC, \DD$
are conjugate and no nonzero solution otherwise.
\end{assumption}

\begin{lemma}\label{lem:sys_calc_f}
Under Assumption~\ref{ass:system}, the calculation of the polynomial $f$ and hence of the matrix $P$ requires $\OO(m^{2\omega})$ operations in $\FF$.
\end{lemma}

\subsection{Second approach: using diagonalization}\label{ss:diagonalisation}

Our strategy may be broken down into the following two steps:  \begin{enumerate}
    \item Since their characteristic polynomial is separable, $U$ and $V$ are diagonalizable over an extension of $\Fq$: we may reduce to the case where they are diagonal.
    \item We find the matrix $P$, which is $R$ multiplied by some element of $\Fq[U]$, by considering its action on some subspaces of $\CC$.
\end{enumerate}

\subsubsection{Reducing to diagonal matrices}
Under the assumption that $U,V$ have a squarefree characteristic polynomial, there is a diagonal matrix $\Delta$
and a matrix $S\in\GL_m(\mathbb{F}_{q'})$ both defined over an extension $\FQ$ of $\Fq$ such that $V=S\Delta S^{-1}$. Such matrices $S, \Delta$ are easily computable.
Therefore, 
\[ U=R^{-1}VR=R^{-1}S\Delta S^{-1}R\] 
and $(\circ)$ is equivalent to
\[ \DD =Sf(\Delta)S^{-1}R \CC R^{-1}Sf(\Delta)^{-1}S^{-1}\] \emph{i.e.},
\[ S^{-1}\DD S=f(\Delta)\cdot S^{-1}R\CC R^{-1}S\cdot
  f(\Delta)^{-1}.\] We may compute bases of $\DD'=S^{-1}\DD S$ and
$\CC'=S^{-1}R\CC R^{-1}S$. The problem at hand is now to compute
$f\in\Fq[t]$ of degree at most $m-1$ such that, given codes
$\CC',\DD'\subset \Fqmm$ of dimension $k$ and a diagonal matrix
$\Delta$, \[ \DD'=f(\Delta)\CC'f(\Delta)^{-1}.\] The reduction is
summed up in the algorithm below.

\begin{algorithm}[H]\label{alg:reducediag}
\SetAlgoLined  
\caption{\textsc{Reducing to diagonal matrices}}
\KwData{Codes $\CC,\DD\subset\Fqmm$\\ Matrices $U,V\in\Fqmm$ with separable characteristic polynomial s.t. $h(\CC)=\Fq\cdot U$ and $h(\DD)=\Fq\cdot V$\\
  Matrix $R\in\GL_m(\Fq)$ such that $V=RUR^{-1}$}
\KwResult{Tuple $(\CC',\DD',\Delta, S)$ where:\\
  $S \in \GL_m(\FF_{q'})$ for some extension $\FF_{q'}/\Fq$,\\ $\Delta \in \GL_m(\FF_{q'})$ diagonal,
  $V=S\Delta S^{-1}$, $\CC'=S^{-1}R \CC R^{-1}S$ and $\DD'=S^{-1}\DD S$.
}
\hrulefill \\
Compute field extension $\FQ$ of $\Fq$ over which $U$ is diagonalizable\\
Compute $S\in\GL_m(\FQ)$ and diagonal $\Delta\in\FQ^{m\times m}$ s.t. $V=S\Delta S^{-1}$\\
Compute bases of $\CC'=S^{-1}R \CC R^{-1}S$ and $\DD'=S^{-1}\DD S$\\
\Return $(\CC',\DD',\Delta,S)$
\end{algorithm}

\subsubsection{Conjugating by the right matrix}

Replacing $q,\CC,\DD$ with $q',\CC',\DD'$, we are now left with the following problem. We are given codes $\CC,\DD\subset \Fqmm$ of dimension $k$ and a diagonal matrix $\Delta$, and need to find a polynomial $f\in\Fq[t]$ such that $\deg(f)<m$ and $\DD=f(\Delta)\CC f(\Delta)^{-1}$. Note that if a polynomial $f$ verifies this, any scalar multiple of $f$ does, so we may assume that $f$ is monic. We may write
\[ \Delta=\begin{pmatrix}
\delta_1 & & & \\
&\delta_2 && \\
&& \ddots &\\
&&& \delta_m
\end{pmatrix}\]
and since $\Delta$ is diagonal, \[f(\Delta)=\begin{pmatrix}
f(\delta_1) & & & \\
&f(\delta_2) && \\
&& \ddots &\\
&&& f(\delta_m)
\end{pmatrix}.\] Our strategy is the following:
\begin{itemize}
    \item Find the entries of $f(\Delta)$.
    \item Knowing $\Delta$, retrieve $f$ using Lagrange interpolation.
\end{itemize}
We may easily find a set $\Lambda$ of $k-1$ non-diagonal indexes $(i,j)\in \{1\dots m\}^2$ such that the respective intersections $\CC(\Lambda),\DD(\Lambda)$ of $\CC,\DD$ with the subspace $E_\Lambda$ of $\Fqmm$ defined by the equations $\{x_{i,j}=0\}_{(i,j)\in \Lambda}$ are one-dimensional. 
\begin{lemma} The subspaces $\CC(\Lambda),\DD(\Lambda)$ satisfy\[ f(\Delta)\CC(\Lambda) f(\Delta)^{-1}=\DD(\Lambda).\]
\begin{proof} Since $\Delta$ is diagonal, so is $f(\Delta)$, and conjugating a matrix by $f(\Delta)$ does not change those of its entries which are equal to zero. Hence, \[ f(\Delta)E_\Lambda f(\Delta)^{-1}=E_\Lambda.\]
The result now follows from the equalities below. \begin{align*}
    f(\Delta)\CC(\Lambda) f(\Delta)^{-1}&= f(\Delta)(\CC\cap E_\Lambda) f(\Delta)^{-1}\\ 
    &= (f(\Delta) \CC f(\Delta)^{-1})\cap (f(\Delta)E_\Lambda f(\Delta)^{-1}) \\
    &= (f(\Delta) \CC f(\Delta)^{-1})\cap E_\Lambda \\
    &= \DD(\Lambda).
\end{align*}
\qed
\end{proof}  
\end{lemma}

We now pick a matrix $C=(c_{ij})_{i,j}\in\CC(\Lambda)$, and a matrix $D=(d_{ij})_{i,j}\in \DD(\Lambda)$. After multiplying by an element of $\Fq$, we may suppose that they have the same characteristic polynomial, and solve $Df(\Delta)=f(\Delta)C$. 
This means solving the system of $m(m-1)/2-(k-1)$ equations \[ d_{ij}f(\lambda_j)=f(\lambda_i)c_{ij}
\tag{ $1\leqslant i<j\leqslant m, (i,j)\not\in S$}\]
which yields the $m$ values $f(\delta_1),\dots,f(\delta_m)$ up to a scalar multiple. We then find a polynomial $f$ corresponding to these values using Lagrange interpolation. 

\begin{remark}
Note that in some rare cases, in particular if $k$ is very small, the matrix $C$ could have too many zeros for the system to determine $f$ uniquely. In that case, picking another set $S$ of coordinates does the trick. 
\end{remark}

\begin{algorithm}[H]\label{alg:findpol}
\SetAlgoLined  
\caption{\textsc{Find the right polynomial}}
\KwData{Codes $\CC,\DD\subset\Fqmm$\\
  Diagonal
  $\Delta = \text{Diag}(\delta_1, \dots, \delta_n)\in\Fqmm$ s.t. $\delta \in \CC$
and $\Delta \in \DD$}
  \KwResult{Polynomial
  $f\in\Fq[t]$ such that $\DD=f(\Delta)\CC f(\Delta)^{-1}$ (if
  exists)}
\hrulefill \\
\While{{\rm true}}{
  Pick set $\Lambda$ of $k-1$ random non diagonal indexes $(i,j)\in \{1,\dots ,m\}^2$ \\
  Compute $\CC(\Lambda)=\CC \cap \{ x_{ij}=0 \}_{(i,j)\in \Lambda}$, $\DD(\Lambda)=\DD \cap \{ x_{ij}=0 \}_{(i,j)\in \Lambda}$\\
  \If{$\dim \CC(\Lambda)=\dim\DD(\Lambda)=1$}{
    Pick $C\in\CC(\Lambda),D\in\DD(\Lambda)$ with the same characteristic polynomial (if exist) \\
    Solve system $d_{ij} u_j=u_i c_{ij}$ for $(i,j)\in \{ 1\dots m\}^2$\\
    \If{The system has no solution}{\Return $\bot$}
    Compute polynomial $f$ such that $f(\delta_{i})=u_i$\\
    \Return $f$}}
\end{algorithm}

\begin{algorithm}[H]\label{alg:findP}
\SetAlgoLined  
\caption{\textsc{Find the right $P$}}
\KwData{Codes $\CC,\DD\subset\Fqmm$\\ Matrices $U,V\in\Fqmm$ with separable characteristic polynomial s.t. $h(\CC)=\Fq\cdot U$ and $h(\DD)=\Fq\cdot V$\\
Matrix $R\in\GL_m(\Fq)$ such that $V=RUR^{-1}$}
\KwResult{A matrix $P$ (if exists), such that $\DD = P \CC P^{-1}$}
\hrulefill \\
Compute $(\CC', \DD', \Delta, S)$ using Algorithm~\ref{alg:reducediag} with inputs
($\CC, \DD, U, V, R$)\\
Compute $f $ using Algorithm~\ref{alg:findpol} with inputs ($\CC',\DD',\Delta$)\\
\If{$f = \bot$}{
  \Return $\bot$}
\Return $S f(\Delta) S^{-1} R$
\end{algorithm}

\subsubsection{Complexity analysis}

\begin{lemma}\label{lem:comp_findP} Given conjugate codes $\CC,\DD\subset \Fqmm$ with
  one-dimensional hulls and generators of these hulls with separable
  characteristic polynomials, the average complexity of finding
  $P\in\Fqmm$ such that $\DD=P\CC P^{-1}$ using Algorithm~\ref{alg:findP} is
  \[\tilde \OO(k^{\omega-1}m^{2} q^{3\sqrt{m}/(\sqrt{\log m}\log q)}).\]
\end{lemma}

\begin{proof}
The smallest field extension $\FQ$ over which the matrix $U$ is diagonalizable is the splitting field of its characteristic polynomial, which has degree $m$. The average degree $d$ of the splitting field of a monic polynomial of degree $m$ over $\Fq$ verifies \cite[Thm.~2]{degree}
\[ d=\exp\left( C\sqrt{m/\log(m)}+\OO\left(\sqrt{m}\log(\log m)/\log(m)
    \right)\right)\] where $C< 3$. This shows that
$d =\OO(q^{3\sqrt{m}/(\sqrt{\log m}\log q)})$. We can do
all the computations over $\FF_{q^d}$, which means the number of
$\Fq$-operations will be that of $\FF_{q^d}$-operations multiplied by
$\tilde\OO(d)$ (using FFT-based algorithm for polynomial arithmetic,
see for instance \cite[Thm.~8.23]{vzg}). Diagonalizing $U$ and $V$ is
done in time $\OO(m^\omega)$. Computing the subspaces $\CC\cap E_\Lambda$,
$\DD\cap E_\Lambda$ is just linear algebra and requires $\OO(k^{\omega-1}m^{2})$
operations in $\FF_{q^d}$. Solving the system of equations takes $\OO(m)$
multiplications in $\FF_{q^d}$. In total, the complexity is
$\tilde\OO(dk^{\omega-1}m^{2})$. \qed
\end{proof}

\section{Recovering $Q$ once we know $P$}\label{sec:recoverQ}

Note first that, given a code $\CC_A$, the probability that a random
code $\DD$ is a conjugate of $\CC_A$ is less than
\[|\GLmq|/|\Gr_k(\Fqmm)|\sim q^{m^2-k(m^2-k)}.\] This is less than
$q^{-m^2}$ for any $m\geqslant 3$ and $2 \leqslant k\leqslant
m^2-2$. Thus, it is highly unlikely that we find matrices $A,B,P$ such
that $\DD_B=P\CC_A P^{-1}$ without there existing a matrix $Q$ such
that $\DD=P\CC Q^{-1}$, and on average, the first $(A,B,P)$ found will
be correct.

The problem we are now trying to solve is the following: given two $k$-dimen\-sional codes $\CC,\DD\subset \Fqmn$, find a matrix $Q\in\GLmq$ such that $\DD=\CC Q$.
Let $(C_1,\dots,C_k)$ be a basis of $\CC$. Given any invertible matrix $Q\in\Fqnn$ such that $C_1Q,\dots,C_kQ\in\DD$, we have $\CC Q=\DD$. Define the linear map \begin{align*}
\psi_\CC\colon \Fqnn &\longrightarrow (\Fqmn)^k\\
Q&\longmapsto (C_1Q,\dots,C_kQ).
\end{align*}
The suitable matrices $Q$ are exactly the elements of
$\psi_\CC^{-1}(\DD^k)\cap\GL_n(\Fq)$. Concretely, computing the space
$\psi_\CC^{-1}(\DD^k)$ requires $\OO(n^2\cdot(mnk)^{\omega-1})$ operations in
$\Fq$. Then, an invertible matrix $Q$ is generally found quite easily
by picking a random element in this space. Note that it may happen
that invertible elements are rare in such a space. However, we claim
that this situation is rather unlikely to happen. Moreover, even in
the worst cases, the problem of finding such a $Q$ can be done in
polynomial time as explained in \cite[Thm.~3.7]{BWB19} and
\cite{CDG20}.

\bibliographystyle{splncs04}
\bibliography{3tensor.bib}

\appendix

\section{Normalizing matrices and characteristic polynomials}\label{sec:normalize}

We consider the map introduced in \Cref{sec:distrib} \begin{align*} \{ U\in \Fqmm\mid \Tr(U)=\Tr(U^2)=0\}/\Fq^\times &\longrightarrow (\Fq^{m-2}-(\Fq^{m-3}\times\{0\}))/\Fq^\times\\
U&\longmapsto (a_{m-3},\dots,a_0) \end{align*}
where the characteristic polynomial of $U$ is $t^m+a_{m-3}t^{m-3}+\dots +a_1t+a_0$ and $\Fq^\times$ acts on $\Fq^{m-2}$ via $\lambda\diamond (a_{m-3},\dots,a_0)=(\lambda^3a_{m-3},\dots,\lambda^m a_0)$.

\begin{lemma}\label{lem:projpond} The set $(\Fq^{m-2}-(\Fq^{m-3}\times\{0\}))/\Fq^\times$ has \[q+q^2+\dots +q^{m-3}\sim q^{m-3}\] elements. 
\end{lemma}

\begin{proof}
The set $(\Fq^{m-2}/\Fq^\times)-\{ (0,\dots,0)\}$ is a subset of the set of $\Fq$-rational points of the weighted projective space $\PP^{m-3}_{3,\dots,m}$ of dimension $m-3$ and weights $3,\dots,m$ over $\Fq$ \cite[Lem.~6]{marcmylove}. By \cite[Lem.~7]{marcmylove}, this has $(q^{m-2}-q)/(q-1)$ elements. \qed
\end{proof}

\begin{lemma}\label{lem:nbpoints} The subset $\Sep\subset (\Fq^{m-2}-(\Fq^{m-3}\times\{0\}))/\Fq^\times$ of classes of separable polynomials has $\sim q^{m-3}$ elements.
\end{lemma}

\begin{proof} The monic inseparable polynomials of degree $m$ over $\Fq$ are the points of an open subset of a hypersurface of degree $2m-2$ in $\PP^{m}$ \cite[§1]{lopez}. The set of inseparable polynomials whose coefficients of degree $m-1,m-2$ vanish is the intersection of this with two hyperplanes that do not contain it. Hence, it is an open subset of a hypersurface of degree $\leq 2m-2$ in $\PP^{m-2}$, and has \[\OO\left((2m-2)q^{m-3}\right)\] elements by the Serre bound \cite[Théorème]{serre}. Since every element of  $\PP^{m-3}_{3,\dots,m}(\Fq)$ has exactly $q-1$ preimages in $\Fq^{m-2}$ \cite[Lem.~7]{marcmylove}, this means that there are $\OO(mq^{m-4})$ classes of inseparable polynomials in $\PP^{m-3}_{3,\dots,m}(\Fq)$. Hence, by \Cref{lem:projpond}, $\Sep$ has $q^{m-3}-\OO(mq^{m-4})\sim q^{m-3}$ elements.\qed
\end{proof}

Here is how to choose and compute a normalized representative of any element $\chi=(a_{m-3},\dots,a_0)\in\Fq^{m-2}$ modulo $\Fq^\times$.  First, the normalized representative of $0$ is itself. Now, consider $\chi\in \Fq^{m-2}-\{ 0\}$. Denote by $i_0<i_1<\dots <i_\ell$ the indices such that $a_{m-i_j}\neq 0$. Choose a generator $g$ of $\Fq^\times$, and write $a_{m-i_j}=g^{s_j}$.  \begin{itemize}
    \item If $i_0$ is prime to $q-1$, there is a unique $\lambda\in\Fq^\times$ such that $\lambda^{i_0} a_{m-i_0}=1$; this $\lambda$ is $g^{-s_0\cdot {i_0}^{-1}{\,\rm mod\,} q}$.  In that case, we choose $\chi'=\lambda\diamond \chi$ to be the normalized representative of $\chi$.
    \item If $d_0\eqdef \gcd(i_0,q-1)>1$, there are $d_0$ elements $\lambda$ satisfying this property. Let us describe how to find the right one.
    \begin{enumerate}
        \item Here is how to compute one such $\lambda$. Write $i_0=d_0i_0'$, and denote by $j_0'$ the inverse of $i_0'$ modulo $q-1$. The set $\Fq^\times/(\Fq^\times)^{i_0}$ has $d_0$ elements: the equivalence classes of $1,g,\dots,g^{d_0-1}$. Compute the Euclidean division $s_0=s_0'\cdot d_0+r_0$ of $s_0$ by $d_0$. Then the element $\lambda_0\eqdef g^{-s_0'j_0'}$ satisfies $\lambda_0^{i_0}a_{m-i_0}=g^{r_0}$. Any product of $\lambda_0$ by a $d_0$-th root of unity in $\Fq$ still satisfies this relation.
        \item Now let $a_{m-i_1}$ be the next nonzero coefficient of $\chi$. We want to normalize $a_{m-i_1}\lambda_0^{i_1}=g^{s_1}$ by multiplying it by a $d_0$-th root of unity. 
        Set $d_1=\gcd(d_0,i_1)$. For any integer $\delta$, denote by $\mu_{\delta}(\Fq)$ the group of $\delta$-th roots of unity in $\Fq$. The set \[\Fq^\times/\mu_{d_0}(\Fq)^{i_1}=\Fq^\times /\mu_{d_0/d_1}(\Fq)\] has $(q-1)d_1/d_0$ elements. Write $i_1=d_1i_1'$ and 
         denote by $j_1'$ the inverse of $i_1'$ modulo $d_0$. Compute the Euclidean division $s_1=s_1'\cdot d_1(q-1)/d_0+r_1$.  The element $\alpha_1=g^{-s_1'j_1'(q-1)/d_0}$ satisfies $\alpha_1^{i_1}g^{s_1}=g^{r_1}$. Set $\lambda_1\eqdef \lambda_0\alpha_1$. We have $\lambda_1^{i_0}a_{m-i_0}=g^{r_0}$ and $\lambda_1^{i_1}a_{m-i_1}=g^{r_1}$.
         \item If $d_1\neq 1$, continue with $d_2=\gcd(d_1,i_2)$. Stop after the $k$-th step if $k=\ell$ or $d_k=1$. Return $\chi'=\lambda_k\diamond \chi$. The algorithm is summed up below. \alain{À revoir}
    \end{enumerate}
\end{itemize}

The element $\chi'\in\Fq^{m-2}$ returned by this algorithm is equivalent to $\chi$. More generally, given equivalent inputs in $\Fq^{m-2}$, it returns the same output.

\begin{algorithm}[H]\label{alg:normalize}
\SetAlgoLined  
\caption{\textsc{Normalize}}
\KwData{Matrix $U\in\Fqmm$, tuple $\chi=(a_{m-3},\dots,a_0)\in\Fq^{m-2}-\{ 0\}$\\
Generator $g$ of $\Fq^\times$}
\KwResult{Matrix $U'\in\Fqmm$, tuple $\chi'\in\Fq^{m-2}-\{ 0\}$}
\hrulefill \\
Set $d=q-1$, $i=2$ and $\lambda=1$\\
\While{$d\neq 1$ \textbf{and} $(a_{m-i-1},\dots,a_1)\neq (0,\dots,0)$}{
Set $i=\min\{ j\in \{i+1\dots m\}\mid a_{m-j}\neq 0\}$\\
Parse $\lambda^i a_{m-i}=g^s$\\
Set $d'=\gcd(d,i)$ and $i'=i/d'$\\
Compute inverse $j'\in\ZZ$ of $i'$ modulo $d$\\
Compute Euclidean division $s=s'\cdot (q-1)d'/d+r$\\
Set $\lambda\leftarrow\lambda g^{-s'j'(q-1)/d}$\\
Set $d\leftarrow d'$, $i\leftarrow i+1$
}
\Return {$\lambda U$, $(\lambda^3 a_{m_3}, \dots, \lambda^m a_0)$}
\end{algorithm}

\begin{remark}\label{rk:precompute}
The complexity of Algorithm \ref{alg:normalize} is ${\OO}(m\log(q)^2)$. To reduce its impact on the complexity of our attack, one can precompute the normalization of the most frequent characteristic polynomials. For instance, on may compute in advance the suitable $\lambda$ for vectors $(a_{m-3},\dots,a_0)$ such that $a_{m-3},a_{m-4}\neq 0$. Since 3 and 4 are coprime, a unique $\lambda$ is found knowing only $a_{m-3},a_{m-4}$: constructing a dictionary $\{(a_{m-3},a_{m-4}):\lambda\}$ allows to precompute the normalization for $(q-1)^2q^{m-4}\sim q^{m-2}$ elements of $\Fq^{m-2}$, that is, almost all of them.
\end{remark}

\section{Proportion of codes with one-dimensional hull}\label{sec:hulldim}

This section explains how the following proposition can be deduced from a similar statement found in the literature.
\hulldimprop*

Sendrier's work \cite{dimhull} gives detailed results about the number
of codes with a hull of given dimension. His results are proven in the
case of codes inside $\Fq^n$ with the usual inner product. In our case
however, we consider $\Fqmm$ endowed with the bilinear form
$(X,Y)\mapsto \Tr(XY)$. Denoting by $\sigma_{n,i}$ the number of
totally isotropic $[n,k]_q$-codes for a given bilinear form, the
number $A_{n,k,1}$-codes whose intersection with their orthogonal
complement has dimension 1 is equal to \cite[Theorem 2]{dimhull}
\[ A_{n,k,1}=\sum_{i=1}^k
  \qbin{n-2i}{k-i}\qbin{i}{1}(-1)^{i-1}q^{(i-1)(i-2)/2}\sigma_{n,i}\]
where $\qbin{n}{k}$ is the \emph{Gaussian binomial coefficient} which denotes
the number of $k$-dimensional linear subspaces of $\Fq^n$.  The proof
of this result does not involve the nature of the considered non-degenerate bilinear form. Sendrier goes on to show \cite[Theorem 3]{dimhull}, using
asymptotic results based on explicit values of $\sigma_{n,i}$ specific
to a bilinear form of discriminant 1, that for $1\leq k\leq
n/2$,\[ A_{n,k,1}q^{k(k+1)/2}=\left(
    \qbin{n}{k}\qbin{k}{1}\prod_{i=0}^{k-1}(q^i-(i\mod
    2))\right)\left( 1+\OO\left(\frac{k}{q^{n/2-1}}\right)\right).\]
There are different formulas of $\sigma_{n,i}$ given in \cite[Theorem
1]{dimhull} depending on the remainders of $n,q$ modulo 2 and 4 and on
the size of $k$. However, they are asymptotically equivalent, which
yields this uniform result. For a bilinear form of different
discriminant, these formulas are simply permuted; a general expression
may be found in \cite[IV, Proposition 3.5]{caldero}. This does not
change the asymptotic result above.  Moreover, for $k\geqslant n/2$,
the number of $[n,k]$-codes with one-dimensional hull is that of
$[n,n-k]$-codes with one-dimensional hull, since the hull of a code
$\CC$ is exactly that of its dual.  In particular, this means that for
any $k$ such that $1\leq k\leq n-1$,
\[ \frac{A_{n,k,1}}{\qbin{n}{k}}= \frac{1}{q}\left( 1+\OO\left(
      \frac{\min(k,n-k)}{q^{n/2-1}}\right)\right)\] or equivalently,
that the proportion of codes whose hull has dimension 1 is
asymptotically equivalent to $1/q$.

\end{document}